\documentclass[%
reprint, superscriptaddress,amsmath,amssymb,aps,
prl,prstper,floatfix,showpacs]{revtex4-1}

\usepackage[FIGTOPCAP,tight]{subfigure}
\usepackage{graphicx}
\usepackage{dcolumn}
\usepackage{bm}

\usepackage{xcolor}
\newcommand{\bra}[1]{\left< #1 \right\vert}
\newcommand{\ket}[1]{\left\vert #1 \right>}
\newcommand{\pare}[1]{\left( #1 \right)}

\newcommand{\cor}[1]{\left[ #1 \right]}

\newcommand{\llav}[1]{\left\lbrace #1 \right\rbrace}

\newcommand{\ave}[1]{\left\langle #1 \right\rangle}

\begin{document}

\preprint{APS/123-QED}

\title{Highly-efficient noise-assisted energy transport in classical oscillator systems}

\author{R. de J. Le\'on-Montiel}
\affiliation{ICFO - Institut de Ciencies Fotoniques, Mediterranean Technology Park, 08860 Castelldefels (Barcelona), Spain}
\author{Juan P. Torres}
\affiliation{ICFO - Institut de Ciencies Fotoniques, Mediterranean Technology Park, 08860 Castelldefels (Barcelona), Spain}
\affiliation{Department of Signal Theory and Communications, Campus Nord D3, Universitat Politecnica de Catalunya, 08034
Barcelona, Spain}


\begin{abstract}
Photosynthesis is a biological process that involves the
highly-efficient transport of energy captured from the sun to a
reaction center, where conversion into useful biochemical energy
takes place. Using a quantum description, Rebentrost \emph{et al.} [New J. Phys. \textbf{11}, 033003
(2009)], and Plenio and Huelga [New J. Phys. \textbf{10}, 113019
(2008)] have explained this high efficiency as the result of the interplay between the quantum
coherent evolution of the photosynthetic system and noise
introduced by its surrounding environment. Even though one can always use a quantum perspective
to describe any physical process, since everything follows the
laws of Quantum Mechanics, is the use of quantum theory imperative
to explain this high efficiency?  Recently, it has been shown by Eisfeld and Briggs [Phys. Rev. E \textbf{85}, 046118 (2012)] that a purely classical model can be used to explain main aspects of the energy transfer in
photosynthetic systems. Using this approach, we demonstrate here
explicitly that highly-efficient noise-assisted energy transport can be found as well in
purely classical systems. The wider scope of
applicability of the enhancement of energy transfer assisted by noise might
open new ways for developing new technologies aimed at enhancing
the efficiency of a myriad of energy transfer systems, from
information channels in micro-electronic circuits to long-distance
high-voltage electrical lines.

\end{abstract}
\maketitle

Because of its undoubted importance for all life on earth,
molecular mechanisms of energy transport in photosynthetic
light-harvesting complexes have been a subject of study for
decades
\cite{frenkel_1931,franck_1938,forster_1965,blankenship2002}. In
recent years, a renewed interest on this topic has arisen
\cite{ball_2011,lambert2013}, mainly due to the unexpected
observation of long-lived electronic coherences in the energy
transfer process of photosynthetic systems, particularly in the
nowadays most widely investigated system, the Fenna-Matthews-Olson
(FMO) complex \cite{engel_2007,engel_2010,scholes_2010}.

As a consequence of these findings, several theoretical studies
have been devoted to describing how coherence effects in a quantum
scenario might play an important role in the remarkably high
efficiency of energy transfer in photosynthetic
systems \cite{ishizaki_2009, whaley_2010, alexandra_2010}. This is
specially notable since it takes place in a scenario apparently
not propitious for the observation of quantum effects. In
particular, it has been suggested that high efficiency transport
arises as a result of the dynamical interplay between the quantum
coherent evolution of the photosynthetic system, and the dephasing
noise introduced by its surrounding environment, a phenomenon
called environment-assisted quantum transport
(ENAQT) \cite{aspuru_2009} or dephasing-assisted energy
transport \cite{plenio_2008}.

As stated in Ref. \cite{aspuru_2012}, ENAQT can be understood as
the suppression of coherent quantum localization mediated by
noise, helping the excitation to move faster through the
photosynthetic system, thus increasing the efficiency of energy
transport. In this way, ENAQT might be seen as a phenomenon that
exists only in a regime where the quantum and classical worlds
overlap. Notwithstanding, making use of the quantum-classical
correspondence of electronic energy transfer presented in Ref.
\cite{eisfeld_2012}, we show here that the same effect can also be
found in purely classical systems. Our departure point is based on
the consideration that aggregates of coupled monomers (such as the
FMO complex) can also be described as a system of
weakly-interacting classical oscillators \cite{briggs_2011}.
We then demonstrate that the noise-assisted enhancement of
transport efficiency in the FMO complex, shown in Refs. \cite{aspuru_2009, plenio_2008}, and based on a
pure quantum formalism, it can also be found in a purely classical
model, without the need to resorting to quantum effects.

For the sake of comparison and clarity, we will first model the
FMO complex as a quantum system of \emph{N} interacting sites,
where the interaction of each site with its surrounding
environment is modeled by a pure dephasing process. We have
adopted this model because of its extended use for describing
noise-assisted energy transfer processes in photosynthetic
systems \cite{aspuru_2009,plenio_2008}. Next, we will present the
classical model of Refs. \cite{briggs_2011,eisfeld_2012}, which corresponds to a system of $N$
weakly-coupled harmonic classical oscillators. In this case,
environmental effects are introduced by assuming that the
frequency of each oscillator vary stochastically as a Gaussian
Markov process. Finally, we will solve both models using the site
energies and coupling coefficients for the FMO complex
of \emph{Prosthecochloris aestuarii} to show that the same
environment-assisted energy transfer effect can be found in both,
the classical and quantum models.

The Hamiltonian of a system comprising $N$ interacting sites in
the presence of a single excitation is given by
\begin{equation}\label{quantum_Ham}
\hat{H}_{S} = \sum_{n=1}^{N} \epsilon_{n}\ket{n}\bra{n} +
\sum_{n\neq m}^{N} V_{nm} \ket{n}\bra{m},
\end{equation}
where $\ket{n}$ denotes the excitation being at site $n$. The
$n$th-site energies and the coupling between sites $n$ and $m$ are
described by $\epsilon_{n}$ and $V_{nm}$, respectively.

We make use of a simple model where the dynamics of the system interacting with a surrounding
environment is described by a Lindblad master equation, which in
the Born-Markov and secular approximations writes as \cite{book_open}
\begin{equation}\label{quantum_evo}
\frac{\partial \hat{\rho}_{nm}}{\partial t} =
-\frac{i}{\hbar}\cor{\hat{H}_{S},\hat{\rho}}_{nm} +
\hat{\mathcal{L}}_{deph}\cor{\hat{\rho}}_{nm} +
\hat{D}\cor{\hat{\rho}}_{nm} .
\end{equation}
Here, the interaction of the system with the environment is
characterized by a pure dephasing process given by the Lindblad
operator $\hat{\mathcal{L}}_{deph}\cor{\hat{\rho}}_{nm} =
-\cor{1/2\,\pare{\gamma_{n} + \gamma_{m}} -
\sqrt{\gamma_{n}\gamma_{m}}\delta_{nm}}\hat{\rho}_{nm},$ with
$\gamma_{n}$ being the dephasing rates. Although the pure dephasing model is not able
to capture important aspects of electronic energy transfer, such as phonon relaxation \cite{ishizaki2_2009},
it provides a useful description of environmental effects in a simple way.
To quantify the transfer of energy from a chosen site $k$ to the \emph{reaction center}, we
have phenomenologically introduced an irreversible decay process (with
rate $\Gamma$) described by the operator $\hat{D}$, which is given
by \cite{book_scully} $\hat{D}\cor{\hat{\rho}}_{nm} =
-\Gamma\llav{\ket{k}\bra{k},\hat{\rho}}_{nm}$, where $\llav{...}$
stands for the anticommutator.

Making use of equation (\ref{quantum_evo}), one can define a measure for
the efficiency of energy transport as the population transferred to
the reaction center, within a time $t$, as
\begin{equation}\label{Q_eff}
Q_{eff} = 2\Gamma \int _{0}^{t} \bra{k}\hat{\rho}\pare{s}\ket{k}
ds .
\end{equation}
Equations (\ref{quantum_evo}) and (\ref{Q_eff}) constitute the
\emph{quantum equations}, which have to be compared with
the equations that will be obtained in the classical model.

For the classical case, we consider an ensemble of $N$ coupled
harmonic oscillators, each with mass $M$ and frequency
$\omega_{n}$. The temporal evolution of the system is described by
a classical Hamiltonian, which in terms of the position $q_{n}$
and momentum $p_{n}$ of each oscillator, reads as
\begin{equation}\label{class_Ham}
H_{S} =  \sum_{n} \pare{\frac{p_{n}^{2}}{2M} +
\frac{M\omega_{n}^{2}}{2}q_{n}^{2}} + \frac{1}{2}\sum_{n\neq
m}K_{nm}q_{n}q_{m},
\end{equation}
where $K_{nm}$ stands for the coupling coefficient between the
oscillators. By defining a new dimensionless complex
amplitude \cite{strocchi_1966}:
$\widetilde{z}_{n}\pare{t} = \widetilde{q}_{n}\pare{t} +
i\widetilde{p}_{n}\pare{t}$, with $\widetilde{q}_{n} =
(M\omega_{n}/2\hbar)^{1/2} q_{n}$ and $\widetilde{p}_{n}=(2\hbar
M\omega_{n})^{-1/2} p_{n}$, the Hamilton equations of motion of
the system can be cast into a single equation
\begin{equation}\label{class_eq}
\frac{\partial \widetilde{z}_{n}}{\partial t} =
-i\omega_{n}\widetilde{z}_{n} -
i\sum_{m}\widetilde{K}_{nm}\mathrm{Re}\llav{\widetilde{z}_{m}}.
\end{equation}
$\mathrm{Re}\llav{...}$ stands for the real part of a complex
number, and $\widetilde{K}_{nm} =
K_{nm}/(M\sqrt{\omega_{n}\omega_{m}})$.

To include environmental effects, we proceed in the same manner as
in the construction of a Kubo oscillator \cite{kubo_1963,fox_1978}.
For this, we assume that the frequency of each classical
oscillator varies randomly as a stochastic process:
$\omega_{n}\pare{t} = \omega_{n} + \phi_{n}\pare{t}$. $\omega_{n}$
is now the average frequency of the $n$th oscillator and
$\phi_{n}\pare{t}$ describes a Gaussian Markov process with zero
average (Wiener process), i.e., $\ave{\phi_{n}\pare{t}} = 0$  and
$\ave{\phi_{n}\pare{t}\phi_{m}\pare{t'}} =
\gamma_{n}\delta_{nm}\delta\pare{t-t'}$, where $\ave{...}$ denotes
stochastic averaging.

\begin{figure}\label{QUANTUM_ENAQT}
\begin{center}
       \includegraphics[width=8cm]{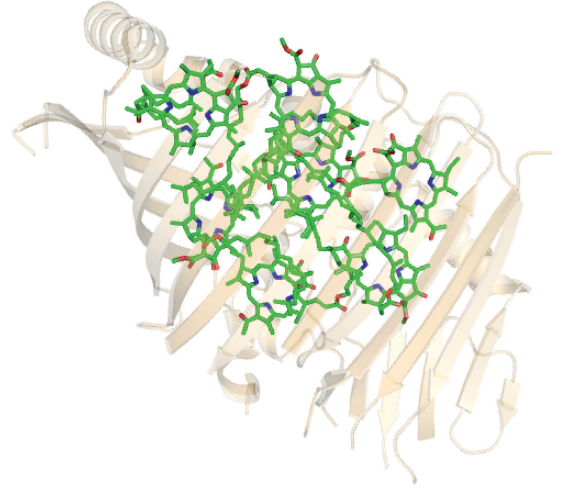}
\end{center}
\label{figure1} \caption{Arrangement of the BChla molecules of a
single unit of the Fenna-Matthews-Olson (FMO) complex. The figure
was created using PyMOL \cite{pymol}, and is based on the PDB entry 3ENI.}
\end{figure}

In Ref. \cite{eisfeld_2012}, it has been shown that one can
transform Eq. (\ref{class_eq}), within the framework of It\^{o}
calculus \cite{kampen_1981}, into a classical master equation that
describes the temporal dynamics of the system of
coupled harmonic oscillators, when interaction with the
surrounding environment is taken into account. To describe the
transfer of excitation from the $k$th oscillator to the reaction
center, we extend this result, and introduce an
irreversible decay process (with rate $\Gamma$), described by
$\mathcal{D}\cor{\sigma}_{nm} =
-\Gamma\llav{\ket{k}\bra{k},\sigma}_{nm}$, where
$\sigma_{nm} = \ave{\widetilde{z}_{n}\widetilde{z}_{m}^{*}}$.
In this way, we can write the \emph{classical master equation} as
\begin{equation}\label{classical_evo_final}
\begin{split}
\frac{\partial \sigma_{nm}}{\partial t} = \mathcal{H}&
\cor{\sigma}_{nm} + \mathcal{L}\cor{\sigma}_{nm} +
\mathcal{D}\cor{\sigma}_{nm} \\
&+
\frac{i}{\hbar}\sum_{j}\pare{V_{mj}\ave{\widetilde{z}_{j}\widetilde{z}_{n}}
- V_{nj}\ave{\widetilde{z}_{j}^{*}\widetilde{z}_{m}^{*}}} ,
\end{split}
\end{equation}
with $V_{nm}  = \widetilde{K}_{nm} \hbar /2$, and
\begin{equation}
\begin{split}
\mathcal{H}\cor{\sigma}_{nm} = -&i\pare{\omega_{n} -
\omega_{m}}\sigma_{nm} \\
&-\frac{i}{\hbar}\sum_{j}\pare{V_{nj}\sigma_{jm} -
V_{jm}\sigma_{nj}},
\end{split}
\end{equation}
\begin{equation}
\mathcal{L}\cor{\sigma}_{nm} =
-\cor{\frac{1}{2}\pare{\gamma_{n}+\gamma_{m}} -
\sqrt{\gamma_{n}\gamma_{m}}\delta_{nm}}\sigma_{nm}.
\end{equation}
The energy transfer efficiency within the ensemble of oscillators is then given by
\begin{equation}\label{C_eff}
C_{eff} = 2\Gamma \int _{0}^{t} \bar{\sigma}_{kk}\pare{s} ds.
\end{equation}
$\bar{\sigma}\pare{t} = \sigma\pare{t}/\sum_{n}\sigma_{nn}$ is the normalized \emph{classical density operator}.

\begin{figure}[t!]\label{QUANTUM_ENAQT}
\begin{center}
       \includegraphics[width=9cm]{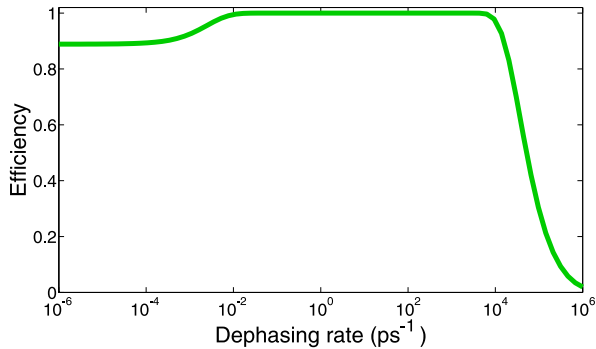}
\end{center}
\label{figure1} \caption{Energy transfer efficiency as a function
of the dephasing rate $\gamma$ obtained from the quantum-mechanical equations.}
\end{figure}

Equations (\ref{classical_evo_final}) and (\ref{C_eff}) represent
the \emph{classical equations}, whose results have to be
compared with their quantum-mechanical counterpart, equations
(\ref{quantum_evo}) and (\ref{Q_eff}). To this end, we can make
use of the site energies and coupling coefficients for
the FMO complex of \textit{P. aestuarii} \cite{renger_2006}. The
FMO is a pigment-protein complex that guides the energy from the
light-harvesting chlorosomes to the reaction center in green
sulfur bacteria \cite{fmo_1,fmo_2}. It is a trimer of three
identical subunits interacting weakly with each other. Each
subunit is composed of seven bacteriochlorophyll-a (BChla)
molecules embedded in a scaffolding of protein molecules, as shown
in Fig. 1. The FMO complex is generally modeled by a network of
seven different sites, where the dynamics of a single excitation
through the complex is governed by the specific values of the site
energies ($\epsilon_{n}$) and the coupling coefficients
($V_{nm}$). In particular, we will use the values of the site
energies and coupling coefficients for \emph{P. aestuarii}, as
stated in Tables 2 and 4 of reference \cite{renger_2006}. The
initial state of the system corresponds to a single excitation in
site $1$. In the FMO, the BChl 3 is in the vicinity of the
reaction center \cite{renger_2006}. Thus, we take this site ($k=3$)
as the main excitation donor to the reaction center, with a
transfer rate estimated to be $\Gamma = 1$ ps$^{-1}$
 \cite{aspuru_2008}. Furthermore, for the pure dephasing process,
we consider that dephasing rates are the same for all sites
($\gamma = \gamma_{n}$) and that the efficiency of energy transfer
is limited by the finite excitation lifetime ($t \sim 1$ ns).

Figure 2 shows the efficiency of energy transfer as a function
of the dephasing rate $\gamma$ obtained by means of the quantum
equations (\ref{quantum_evo}) and (\ref{Q_eff}). Notice
that at low dephasing, i.e., with environment effects not considered,
coherent evolution of the system leads to an efficiency of
about $90\%$. When increasing the dephasing, efficiency grows to
almost $100\%$, showing that the environment affects the system in
such a way that it becomes more efficient for transferring energy
to the reaction center. Finally, for stronger dephasing,
efficiency drops rapidly and almost no energy is transferred to
the reaction center. Qualitatively similar results have also been obtained
for the case of the FMO complex of \emph{Chlorobium tepidum} \cite{aspuru_2009}.

We now turn our attention to the case of the classical model by solving
equations (\ref{classical_evo_final}) and (\ref{C_eff}).
Figure 3 shows the efficiency of the energy transfer as a function
of the dephasing rate. We observe that the same noise-assisted
effect is also present in the purely classical model. For the sake
of comparison, Fig. 3 also shows the solution of
the quantum mechanical model (dashed line). Notice that both
solutions agree for dephasing rates up to $10^{3}$ ps$^{-1}$.
However, for larger values of dephasing the quantum and classical
solutions differ from each other. This is in agreement with the
fact that both solutions are the same, provided that the condition
$\gamma\ll\omega_{n}$ is satisfied \cite{eisfeld_2012}.

Noise-assisted energy transport in disordered systems has been
understood as the suppression of coherent quantum localization
through noise, bringing the detuned quantum levels into resonance
and thus facilitating the energy
transfer \cite{aspuru_2009,aspuru_2012}.  Notwithstanding, the
results presented here show that the same effect can also be found
in purely classical systems. This implies that one can make use of
such systems in order to simulate the intricate energy transfer
mechanisms that take place in molecular aggregates, such as the
photosynthetic FMO complex.

Recently, it has been suggested that classical LC circuit
oscillators (where L stands for inductance and C for capacitance)
can be used to model coupled quantum two-level
systems \cite{briggs_2012}. Hence, one could devise an experimental
apparatus comprising eight electrical oscillators with the eighth acting as
the reaction center, which would be strongly
coupled to one of the remaining oscillators. Then, by
stochastically modulating the frequencies $\omega_{n}$, and properly
controlling the noise intensity $\gamma_{n}$, one would be able to observe the noise-assisted
energy transfer phenomenon by monitoring the signal present in the eighth oscillator. These
classical simulations could be further used to compare with the recent
experimental proposal of noise-assisted transport based on coupled quantum-optical
cavities \cite{plenio_2011}.

\begin{figure}[t!]\label{CLASSICAL_ENAQT}
\begin{center}
       \includegraphics[width=9cm]{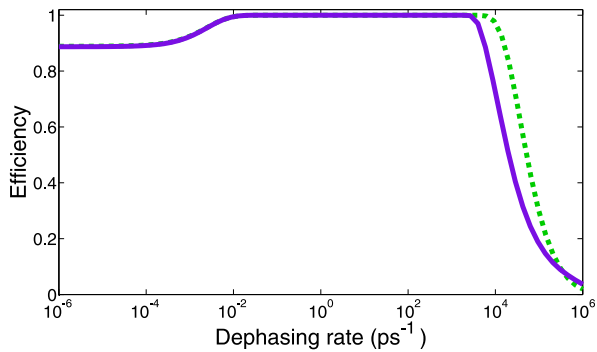}
\end{center}
\label{figure1} \caption{Energy transfer efficiency as a function
of the dephasing rate $\gamma$ obtained from the classical
equations (solid line). For the sake of comparison, we have also
included here the curve shown in Fig. 2, which corresponds to the solution of the quantum equations (dashed line).}
\end{figure}

The concept of noise-assisted energy transport has been extensively used for describing the inner working of quantum and classical systems \cite{hanggi_2009}. Along these lines, the particular enhancement effect described in this paper might open a new research direction towards new methods for enhancing the efficiency of a myriad of
energy transport systems that inevitable live in a noisy
environment, from small-scale information and energy transfer
systems in microwave and photonic circuits, to long-distance
high-voltage electrical lines. In this way, a specific feature initially
conceived in a quantum scenario ({\em environment-assisted energy
transport}) is shown to arise as well in a purely classical
context, widening thus the scope of possible
\textit{quantum-inspired} technological applications.

To conclude, the search and demonstration of systems where to
observe quantum-mechanical effects with no classical counterpart
is a subject of lively interest and debate \cite{zimanyi_2010,miller2012,popescu2012}. Biological systems are not, in
principle, a propitious scenario for the observation of quantum
features, such as quantum superposition, interference or
entanglement. Nevertheless, one can always use a quantum
perspective to describe any physical process, since everything
follows the laws of Quantum Mechanics. This does not mean,
however, that in certain cases a purely classical model may not
similarly reproduce some of the results predicted by the full
quantum-mechanical model, since classical physics emerge, after all, from quantum physics under
many circumstances.

This work was supported by projects FIS2010-14831 and FET-Open grant number: 255914 (PHORBITECH). This
work has also been partially supported by Fundacio Privada Cellex
Barcelona.

\end{document}